\documentclass{PoS}

\usepackage{amsmath,amssymb}

\title{Noncommutative field theories with harmonic term}

\ShortTitle{Noncommutative field theories with harmonic term}

\author{\speaker{Axel de Goursac}%
        \thanks{Work
supported by the Belgian Interuniversity Attraction Pole (IAP) within the framework ``Nonlinear systems, stochastic processes, and statistical mechanics'' (NOSY).}\\
        Universite Catholique de Louvain\\
        E-mail: \email{axelmg@melix.net}}

\abstract{The UV-IR mixing of scalar field theory on the Moyal space is removed by the harmonic term, so that the theory is renormalizable. We will present  different
properties of this scalar model and its associated gauge theory, which is
candidate to renormalizability. The supergeometric interpretation of the
harmonic term, for both scalar and gauge models, and related to the
Langmann-Szabo duality, will be exposed. }

\FullConference{Corfu Summer Institute on Elementary Particles and Physics - Workshop on Non 			Commutative Field Theory and Gravity,\\
		September 8-12, 2010\\
		Corfu Greece}
		
\newcommand\caA{{\mathcal A}}

\newcommand\caF{{\mathcal F}}

\newcommand\caL{{\mathcal L}}

\newcommand\wx{{\widetilde x}}
\newcommand\gone{{ \mathchoice {1\mskip-4mu\mathrm{l} } {1\mskip-4mu\mathrm{l} }{1\mskip-4.5mu\mathrm{l} } {1\mskip-5mu\mathrm{l}} }}
\newcommand\gR{{\mathbb R}}

\newcommand\gC{{\mathbb C}}

\newcommand\gZ{{\mathbb Z}}

\newcommand\algA{{\mathbf A}}

\newcommand\kg{{\mathfrak g}}

\newcommand\kh{{\mathfrak h}}

\newcommand\kX{{\mathfrak X}}

\newcommand\fois{\mathord{\cdot}}
\DeclareMathOperator{\tr}{Tr} 

\newcommand\dd{{\text{\textup{d}}}}

\begin{document}

\section*{Introduction}

In the past few years, there has been a growing interest in the noncommutative quantum field theories (for a review, see \cite{Wulkenhaar:2006si}). These theories on ``spaces'' coming from noncommutative geometry \cite{Connes:1994} are indeed good candidates for new physics behind the Standard Model of particle physics. At the Planck scale, one can wonder wether events are not localizable with an arbitrary precision, what could be modelized by a noncommutative space-time \cite{Doplicher:1994tu}. Noncommutative field theories have also relationship with classical and quantum gravity \cite{Chamseddine:2006ep,Freidel:2006}. Moreover, field theories defined on the Moyal space, one of the simplest example of noncommutative space, can be seen as an effective regime of string theory \cite{Seiberg:1999vs} and matrix theory \cite{Connes:1997cr}. They also may have implications in the description of non-local physics in the presence of a background field, like the quantum Hall effect \cite{Hellerman:2001rj}.

One of the key ingredients of quantum field theory is the renormalization. However, the standard $\phi^4$ real scalar theory and the Yang-Mills theory on the Euclidean Moyal space suffer from a new type of divergence, the Ultraviolet-Infrared (UV-IR) mixing \cite{Minwalla:1999px}, spoiling their renormalizability. Recently, this problem has been solved in the scalar case \cite{Grosse:2004yu} by the addition of a harmonic term in the action. Then, the resulting theory possess new features according to renormalization. Such a change is also possible in the gauge case \cite{deGoursac:2007gq} (see \cite{deGoursac:2009gh} for a review). These theories with harmonic term have later been interpreted in a supergeometric framework \cite{deGoursac:2008bd,deGoursac:2010zb,Bieliavsky:2010su}. Note that another interpretation for the harmonic term has been developed for the scalar theory \cite{Buric:2009ss,Buric:2010xs}. Noncommutative quantum field theory with harmonic term has also been discussed in the Minkowskian setting \cite{Fischer:2008dq,Zahn:2010yt,Fischer:2010zg}.

In this paper which is based on a conference given at the {\it Corfu Summer Institute on Elementary Particles and Physics 2010}, we present the scalar and gauge theories with harmonic term on the Moyal space, and expose their supergeometric interpretation. Section \ref{qft} is devoted to the definition of the Moyal space, and of noncommutative scalar and gauge theories with harmonic term. Then, we introduce in section \ref{lsd} an important symmetry of the scalar field theory with harmonic term, called the Langmann-Szabo duality, and expose its group reformulation. In section \ref{super}, the supergeometric framework, namely the superalgebra $\algA_\theta$, its differential calculus and its theory of connections, is defined and related to field theories with harmonic term. Finally, in section \ref{disc}, the Langmann-Szabo duality is discussed within the supergeometric setting.

\section{QFT with harmonic term on the Moyal space}
\label{qft}

Let us first introduce the Moyal space. It is a deformation quantization (see \cite{Bayen:1978}) of the euclidean space $\gR^4$, which means that the functions $\gR^4\to\gC$ are endowed with a noncommutative product depending on a deformation parameter $\theta$ and coinciding with the usual commutative product for $\theta=0$. By setting the matrix
\begin{equation}
\Sigma=\begin{pmatrix} 0 & -1 & 0 & 0 \\ 1 & 0 & 0 & 0 \\ 0 & 0 & 0 & -1 \\ 0 & 0 & 1 & 0 \end{pmatrix},\label{matrsigma}
\end{equation}
and $\Theta_{\mu\nu}=\theta \Sigma_{\mu\nu}$, the deformed star-product $\star$ satisfies the usual commutation relations:
\begin{equation}
[x_\mu,x_\nu]_\star=i\Theta_{\mu\nu},\label{comrel}
\end{equation}
where $[f,g]_\star=f\star g-g\star f$ is the star-commutator. We also introduce the following notations for $x,y\in\gR^4$:
\begin{equation}
x\wedge y= 2 x_\mu\Theta^{-1}_{\mu\nu}y_\nu,\qquad \wx_\mu=2\Theta^{-1}_{\mu\nu}x_\nu.\label{convmoy}
\end{equation}
Then, the Moyal star-product can be expressed on functions $f,g:\gR^4\to\gC$ by
\begin{equation}
(f\star g)(x)=\frac{1}{\pi^4\theta^4}\int \dd^4y\,\dd^4z\ f(x+y)\,g(x+z)e^{-iy\wedge z}.\label{prodmoy}
\end{equation}
Let us also recall two important properties of this product: the integral of a star-product of two functions corresponds to the integral of the commutative product of these functions, and the derivatives can be expressed with a star-commutator.
\begin{equation}
\int \dd^4x\, (f\star g)(x)=\int\dd^4x\, f(x)g(x),\qquad\partial_\mu f=-\frac i2[\wx_\mu,f]_\star.\label{propmoy}
\end{equation}

\medskip

The extension of the real $\phi^4$ scalar field theory from the commutative euclidean space $\gR^4$ to the Moyal space consists to replace the commutative product in the action by the Moyal one:
\begin{equation*}
S(\phi)=\int \dd^4x\Big(\frac 12\partial_\mu\phi\star\partial_\mu\phi +\frac{m^2}{2}\phi\star\phi +\lambda\phi\star\phi\star\phi\star\phi\Big).
\end{equation*}
By using the first property of \eqref{propmoy}, we see that the action is given by
\begin{equation}
S(\phi)=\int \dd^4x\Big(\frac 12(\partial_\mu\phi)^2 +\frac{m^2}{2}\phi^2 +\lambda\phi\star\phi\star\phi\star\phi\Big),\label{actphi4}
\end{equation}
so that only the vertex term is changed with respect to the commutative theory. Feynman rules can be computed with the expression of the Moyal product \eqref{prodmoy}, but this leads to a non-renormalizable theory. Indeed, this model suffers from a new type of divergence called UV-IR mixing \cite{Minwalla:1999px}, due to the non-locality of the Moyal product. This mixing of scales can be seen at the level of the relation \eqref{comrel} which  is verified by the product.

The first solution to this problem of UV-IR mixing has been discovered by H. Grosse and R. Wulkenhaar by adding a harmonic term to the scalar action. The action becomes:
\begin{equation}
S(\phi)=\int \dd^4x\Big(\frac 12(\partial_\mu\phi)^2 +\frac{\Omega^2}{2}\wx^2\phi^2+\frac{m^2}{2}\phi^2 +\lambda\phi\star\phi\star\phi\star\phi\Big),\label{actharm}
\end{equation}
so that the theory is renormalizable to all orders in perturbation \cite{Grosse:2004yu} when the parameter $\Omega\neq 0$. We recall that $\wx_\mu$ is defined in \eqref{convmoy}. A Connes-Kreimer Hopf algebra encoding its renormalization has been constructed in \cite{Tanasa:2007xa,Tanasa:2009hb}.

This theory has several remarkable properties. At the fixed point of the renormalization group $\Omega=1$, the beta function of the constant $\lambda$ vanishes \cite{Disertori:2006nq}, contrary to the commutative $\phi^4$ theory where a Landau ghost appeared. The vacuum solutions of the theory with a negative mass term ($m^2<0$) have been exhibited in \cite{deGoursac:2007uv}. Moreover, even if the choice of the matrix $\Sigma$ \eqref{matrsigma} breaks the rotational invariance of the model, it can be restored at all orders in the renormalization procedure by considering a family of actions labeled by matrices of the type of $\Sigma$ \cite{deGoursac:2009fm} (this means that $\Sigma$ has to be considered as a tensor). Note also that there exists another solution for the UV-IR mixing problem of the real scalar $\phi^4$-theory \cite{Gurau:2008vd,Tanasa:2010fk}.

\medskip

The setting of gauge theories can also be extended to the context of the Moyal deformation (see \cite{deGoursac:2007gq,Cagnache:2008tz} for more details). We consider here a noncommutative $U(1)$-gauge theory: the space is noncommutative because of the deformed product on functions, but the gauge group $U(1)$ is abelian.

Gauge potentials are real functions $A_\mu:\gR^4\to\gR$, like in the commutative setting, but the product involved in the definition of the field strength $F_{\mu\nu}$ and of the gauge transformations has to be replaced by the Moyal one:
\begin{align*}
F_{\mu\nu}&=\partial_\mu A_\nu-\partial_\nu A_\mu-i[A_\mu,A_\nu]_\star,\\
A_\mu^g&=g\star A_\mu\star g^\dag+ig\star\partial_\mu g^\dag,
\end{align*}
where $g:\gR^4\to\gC$ is a gauge transformation, namely satisfying $g^\dag\star g=g\star g^\dag=1$. Note however that there is a major difference with the commutative setting. Indeed, because of the second property of \eqref{propmoy}, one can construct the following expression, called covariant coordinate:
\begin{equation*}
\caA_\mu=A_\mu+\frac12\wx_\mu,
\end{equation*}
which can be checked to transform covariantly under a gauge transformation: $\caA_\mu^g=A^g_\mu+\frac12\wx_\mu=g\star\caA_\mu\star g^\dag$.

Like in the scalar case, the Yang-Mills action for the Moyal space
\begin{equation*}
S(A)=\int\dd^4x\Big(\frac 14F_{\mu\nu}\star F_{\mu\nu}\Big)
\end{equation*}
suffers also from the UV-IR mixing \cite{Minwalla:1999px}. The analog of the harmonic term for the gauge theory, which has to be gauge-invariant, has been found by computing an effective action from the Grosse-Wulkenhaar model \cite{deGoursac:2007gq,Grosse:2007dm}:
\begin{equation}
S(A)=\int\dd^4x\Big(\frac 14F_{\mu\nu}\star F_{\mu\nu}+\frac{\Omega^2}{4}\{\caA_\mu,\caA_\nu\}_\star^2+\kappa \caA_\mu\star\caA_\mu\Big),\label{acteff}
\end{equation}
where $\{f,g\}_\star =f\star g+g\star f$ is the star-anticommutator, $\Omega$ and $\kappa$ are real parameters, and we recall that $F_{\mu\nu}$ and $\caA_\mu$ directly depend on the gauge potential $A_\mu$. It is therefore a candidate to a renormalizable noncommutative gauge theory. By reexpressing this action in terms of $A_\mu$, one obtains:
\begin{multline*}
S(A)=\int\dd^4x\Big(\frac{\Omega^2}{4}\wx^2\wx_\mu A_\mu+\kappa\wx_\mu A_\mu-\frac 12A_\mu\partial^2 A_\mu+\frac{\Omega^2}{2}\wx^2 A_\mu A_\mu+\kappa A_\mu A_\mu\\
-\frac 12(1-\Omega^2)(\partial_\mu A_\mu)^2+\Omega^2(\wx_\mu A_\mu)^2-i(\partial_\mu A_\nu)[A_\mu,A_\nu]_\star\\
+\Omega^2\wx_\mu A_\nu\{A_\mu,A_\nu\}_\star-\frac 14[A_\mu,A_\nu]_\star^2+\frac{\Omega^2}{4}\{A_\mu,A_\nu\}_\star^2\Big),
\end{multline*}
up to a constant term. We can notice that the harmonic term responsible of the removing of the UV-IR mixing in the scalar case is present in this action. However, the presence of linear terms in the field $A_\mu$ indicates that there is no trivial vacuum solution for this action. Consequently, the solutions of the equation of motion for this model have been exhibited in \cite{deGoursac:2008rb}. Note that the ghost sector of a similar gauge model has been studied in \cite{Blaschke:2007vc,Blaschke:2009aw}.

\section{Langmann-Szabo duality}
\label{lsd}

In this section, we review a special symmetry of the model \eqref{actharm} called the Langmann-Szabo duality which seems to play an important role for its renormalizability, and its group interpretation which will be useful in section \ref{disc}.

The Langmann-Szabo duality can be expressed as a symplectic Fourier transformation on the real field $\phi$:
\begin{equation}
\hat\phi(p)=\frac{1}{(\pi\theta)^2}\int\dd^4x\,\phi(x)e^{\pm i p\wedge x}\label{fouriersympl}
\end{equation}
where $p\wedge x$ is defined in \eqref{convmoy}. Note that the sign in the phase depend on the place of the field $\phi$ in a product (see \cite{deGoursac:2010zb} for more details). Usual properties like the Parseval-Plancherel equality and the exchange of derivative and multiplication by the coordinate under a Fourier transformation are still valid, so that one can show that the action \eqref{actharm} satisfies:
\begin{equation*}
S[\phi;m,\lambda,\Omega]=\Omega^2\ S\Big[\hat\phi,\frac{m}{\Omega},\frac{\lambda}{\Omega^2},\frac{1}{\Omega}\Big].
\end{equation*}
This property is called the Langmann-Szabo covariance \cite{Langmann:2002cc}, and is not verified by the standard $\phi^4$ action \eqref{actphi4}. For $\Omega=1$, the action \eqref{actharm} is invariant under this duality. Thus, this symmetry seems to be related to the renormalization of this theory. However, it is no more a symmetry for the associated gauge theory \eqref{acteff}.

\medskip

Let us now expose the group interpretation of this duality. It has been given in \cite{Bieliavsky:2008qy}. We first introduce the Heisenberg algebra $\kh=\gR^9$ associated to the phase space $\gR^8$ with symplectic structure $\omega=\begin{pmatrix} 0 & -\Sigma \\ -\Sigma & 0 \end{pmatrix}$. Its commutation relations are given by: $\forall x,y,p,q\in\gR^4$, $\forall s,t\in\gR$,
\begin{equation}
[(x,p,s),(y,q,t)]=(0,0,-x_\mu\Sigma_{\mu\nu}q_\nu-p_\mu\Sigma_{\mu\nu}y_\nu),\label{heisenberg}
\end{equation}
where $\Sigma$ is defined in \eqref{matrsigma} on the configuration space $\gR^4$. We also recall the definition of the symplectic group for the phase space:
\begin{equation*}
Sp(\gR^8,\omega)=\{M\in GL(\gR^8),\quad M^T\omega M=\omega\}.
\end{equation*}
Then, we consider the metaplectic representation of the symplectic group, which can be constructed from the Heisenberg algebra (for mathematical details, see \cite{deGoursac:2010zb}):
\begin{equation*}
\mu: Sp(\gR^8,\omega)\to\caL(L^2(\gR^4)).
\end{equation*}
This representation indeed associates to any phase space transformation a field transformation (on the configuration space). Then, we will look at the Langmann-Szabo duality which is a field transformation, at the level of the symplectic group with the representation $\mu$.

For $M=\frac{\theta}{2}\begin{pmatrix} 0 & \frac{4}{\theta^2}\gone \\ \gone & 0 \end{pmatrix}$ in the group $Sp(\gR^8,\omega)$, one can check that $\mu(M)$ is exactly a symplectic Fourier transformation \eqref{fouriersympl}, that is the Langmann-Szabo duality. Let us also consider the infinitesimal generator $Z=\begin{pmatrix} 0 & -\frac{4\Omega^2}{\theta^2}\Sigma \\ -\Sigma & 0 \end{pmatrix}$ in the Lie algebra $\mathfrak{sp}(\gR^8,\omega)$. The image by the infinitesimal metaplectic representation $\dd\mu$ is exactly the operator involved in the quadratic part of the action \eqref{actharm}:
\begin{equation*}
2i\dd\mu(Z)=-\partial^2+\Omega^2\wx^2.
\end{equation*}
Then, the subgroup of $Sp(\gR^8,\omega)$ which leaves covariant the quadratic part of the action \eqref{actharm} with the metaplectic representation corresponds to the subgroup of $Sp(\gR^8,\omega)$ which conserves the form of $Z$ by the adjoint representation. It is therefore easy to see that the symmetries leaving the model \eqref{actharm} covariant by the metaplectic representation are only generated by $M$ and the position space transformations. This shows the particular role of the Langmann-Szabo duality.

Since $Sp(\gR^8,\omega)$ acts on the Heisenberg algebra $\kh$, we compute the action of $M$:
\begin{equation}
M.(x,p,s)=\Big(\frac{2}{\theta}p,\frac{\theta}{2}x,s\Big),\label{actsympl}
\end{equation}
as it will be useful in section \ref{disc}.

\section{Supergeometric interpretation}
\label{super}

We expose in this section the supergeometric interpretation of the scalar theory with harmonic term and its associated gauge theory, introduced in \cite{deGoursac:2008bd,Bieliavsky:2010su}.

By using the second property of \eqref{propmoy} as well as the identity $\wx_\mu\fois f=\frac12\{\wx_\mu,f\}_\star$, one can reexpress the action \eqref{actharm} into:
\begin{equation*}
S(\phi)=\int \dd^4x\Big(-\frac 18[\wx_\mu,\phi]_\star^2 +\frac{\Omega^2}{8}\{\wx_\mu,\phi\}_\star^2+\frac{m^2}{2}\phi^2 +\lambda\phi\star\phi\star\phi\star\phi\Big).
\end{equation*}
In the same way, using the relation between the field strength and the covariant coordinate $F_{\mu\nu}=\Theta^{-1}_{\mu\nu}-i[\caA_\mu,\caA_\nu]_\star$, the gauge theory \eqref{acteff} can also be reexpressed as
\begin{equation}
S(A)=\int\dd^4x\Big(-\frac 14[\caA_\mu,\caA_\nu]_\star^2+\frac{\Omega^2}{4}\{\caA_\mu,\caA_\nu\}_\star^2+\kappa \caA_\mu\star\caA_\mu\Big),\label{actgaugesym}
\end{equation}
up to a constant term. One sees immediately a sort of symmetry between star-commutators and star-anticommutators in both actions. This symmetry is reminiscent of a grading, and it motivates the introduction of a superalgebra in the following.

\medskip

The right superalgebra to consider here is the Moyal-Clifford algebra $\algA_\theta$, which is a deformation quantization of the superspace $\gR^{4|1}$ (see \cite{Bieliavsky:2010su}). Let $\algA_\theta$ be the space of functions from $\gR^{4|1}$ to $\gC$ with a certain regularity (smooth, with all derivatives bounded). For $f\in\algA_\theta$, $x\in\gR^4$ and $\xi\in\gR^{0|1}$, one has the decomposition
\begin{equation*}
f(x,\xi)=f_0(x)+f_1(x)\xi,
\end{equation*}
where we recall that $\xi$ is an anticommuting variable, and $f_i$ are smooth functions $\gR^4\to\gC$. $\algA_\theta$ has a $\gZ_2$-grading given by the power of the variable $\xi$, such that the degree of a homogeneous element $f$ is denoted by $|f|$. The deformed product of $\algA_\theta$ is compatible with the grading and can be expressed in terms of the Moyal product (both are denoted by $\star$): $\forall f,g\in\algA_\theta$,
\begin{equation*}
(f\star g)(x,\xi)= (f_0\star g_0)(x)+\alpha (f_1\star g_1)(x)+ (f_0\star g_1)(x)\xi+(f_1\star g_0)(x)\xi,
\end{equation*}
where $\alpha$ is a second parameter of the deformation of $\gR^{4|1}$ called superization parameter. Indeed, for $\alpha=0$, only the even part $\gR^4$ of $\gR^{4|1}$ is deformed with deformation parameter $\theta$. The unit of the algebra is the function $1$. We also endow $\algA_\theta$ with an involution: $f^\dag(x,\xi)=\overline{f_0(x)}+\overline{f_1(x)}\xi$, and a (non-graded) trace: $\tr(f)=\int\dd^4x\, f_0(x)$. Note that the graded bracket of $\algA_\theta$ is given in terms of the Moyal commutator and the Moyal anticommutator:
\begin{equation}
[f,g]_\star(x,\xi)= [f_0,g_0]_\star(x)+\alpha\{f_1,g_1\}_\star(x)+[f_0,g_1]_\star(x)\xi+[f_1,g_0]_\star(x)\xi. \label{grbracket}
\end{equation}

\medskip

Then, we introduce a differential calculus for the superalgebra $\algA_\theta$ which will allow to obtain the actions \eqref{actharm} and \eqref{acteff}. This differential calculus is based on the graded derivations of $\algA_\theta$ (see \cite{DuboisViolette:1988cr,deGoursac:2008bd}), that is on endomorphisms $\kX$ of $\algA_\theta$ which satisfy $\forall f,g\in\algA_\theta$, $\kX(f\star g)=\kX(f)\star g+(-1)^{|\kX||f|}f\star\kX(g)$. Inner (graded) derivations are given by (graded) brackets: $[f,\fois]_\star$.

Let us consider $\kg$ the smallest graded Lie subalgebra of $\algA_\theta$ containing $-\frac{i}{2}\wx_\mu$ and $-\frac{i}{2}\wx_\mu\xi$ (whose importance will be noticed in the following). One can check with closure relations that $(1,i\xi,-\frac{i}{2}\wx_\mu,-\frac{i}{2}\wx_\mu\xi,\frac{i}{2}\wx_\mu\wx_\nu)$ is a basis of $\kg$. Then, the one-forms of the differential calculus based on the derivations $\kX=[f,\fois]_\star$, for $f\in\kg$, are linear applications $\omega:\kg\to\algA_\theta$ which vanish on $1$. We denote their coefficients by:
\begin{equation}
\omega_0=\omega([i\xi,\fois]_\star),\qquad \omega_\mu=\omega([-\frac{i}{2}\wx_\mu,\fois]_\star),\qquad \omega_{\overline\mu}=\omega([-\frac{i}{2}\wx_\mu\xi,\fois]_\star),\qquad \omega_{(\mu\nu)}=\omega([\frac{i}{2}\wx_\mu\wx_\nu,\fois]_\star).\label{indexform}
\end{equation}
The differential acts on the elements $f\in\algA_\theta$ as
\begin{align}
&(\dd f)_\mu=\partial_\mu f_0+\partial_\mu f_1\xi=\partial_\mu f,\qquad (\dd f)_{\overline\mu}=i\alpha\wx_\mu f_1+\partial_\mu f_0\xi,\nonumber\\
&(\dd f)_0=-2i\alpha f_1,\qquad (\dd f)_{(\mu\nu)}=-\frac 14(\wx_\mu\partial_\nu f_0+\wx_\nu\partial_\mu f_0+\wx_\mu\partial_\nu f_1\xi+\wx_\nu\partial_\mu f_1\xi).\label{differ}
\end{align}
One sees that the derivation associated to $-\frac{i}{2}\wx_\mu$ is therefore corresponding to the usual differential calculus of the Moyal space. $-\frac{i}{2}\wx_\mu\xi$ is its graded counterpart.
\medskip

We can now construct the $\phi^4$ scalar action for this differential calculus. Let us consider the field $(1+\xi)\phi(x)$ on $\gR^{4|1}$ (see section \ref{disc}). We define the scalar action to be
\begin{equation}
S(\phi)=\tr\Big( \frac12\sum_{f} |\dd((1+\xi)\phi)([f,\fois]_\star)|^2+\frac{m^2}{2}((1+\xi)\phi)^2+\lambda (((1+\xi)\phi)\star((1+\xi)\phi))^2\Big),\label{actsuperphi4}
\end{equation}
where the sum on $f$ is over $i\xi,-\frac{i}{2}\wx_\mu,-\frac{i}{2}\wx_\mu\xi$. We will see in section \ref{disc} why $f=\frac{i}{2}\wx_\mu\wx_\nu$ is not considered. The action \eqref{actsuperphi4} can be simplified as
\begin{equation*}
S(\phi)=\int\dd^4x\Big(\frac 12(\partial_\mu\phi)^2 +\frac{\alpha^2}{2}\wx^2\phi^2+(\frac{m^2}{2}+\frac{4\alpha^2}{\theta})\phi^2 +\lambda(1+\alpha)^2\phi\star\phi\star\phi\star\phi\Big),
\end{equation*}
so that this model coincides with the Grosse-Wulkenhaar model \eqref{actharm} which is renormalizable.

\medskip

The theory of connections for this differential calculus leads also to the gauge theory \eqref{acteff} associated to the Grosse-Wulkenhaar model. To each inner derivation of $\kg$, we associate a gauge potential:
\begin{equation*}
-\frac{i}{2}\wx_\mu\to A_\mu^0,\qquad -\frac{i}{2}\wx_\mu\xi\to A_\mu^1\xi,\qquad
i\xi\to \varphi\xi,\qquad \frac{i}{2}\wx_\mu\wx_\nu\to G_{\mu\nu}.
\end{equation*}
The gauge transformations of these potentials are:
\begin{align*}
&(A_\mu^0)^g=g\star A_\mu^0\star g^\dag+ig\star\partial_\mu g^\dag, \qquad
(A_\mu^1)^g=g\star A_\mu^1\star g^\dag+ig\star\partial_\mu g^\dag,\\
&\varphi^g=g\star\varphi\star g^\dag,\qquad
(G_{\mu\nu})^g=g\star G_{\mu\nu}\star g^\dag-\frac i4g\star(\wx_\mu\partial_\nu g^\dag)-\frac i4g\star(\wx_\nu\partial_\mu g^\dag),
\end{align*}
where $g:\gR^4\to\gC$ satisfies $g^\dag\star g=g\star g^\dag=1$. According to the gauge transformations, we can perform the simplification: $\varphi=0$, $G_{\mu\nu}=\frac12\wx_\mu\wx_\nu$, $A_\mu^0=A_\mu^1=:A_\mu$. Then, the (graded-antisymmetric) field strength $\caF$ (or 2-form of curvature) for this differential calculus can be computed:
\begin{align}
&\caF_{\mu\nu}=F_{\mu\nu},\qquad \caF_{\mu\overline\nu}=F_{\mu\nu}\xi,\qquad \caF_{\overline\mu\overline\nu}=-i\alpha\{\caA_\mu,\caA_\nu\}_\star,\nonumber\\
& \caF_{\mu (\nu\rho)}=2\Theta^{-1}_{\nu\mu}\caA_\rho +2\Theta^{-1}_{\rho\mu}\caA_\nu,\qquad \caF_{\overline{\mu} (\nu\rho)}=(2\Theta^{-1}_{\nu\mu}\caA_\rho +2\Theta^{-1}_{\rho\mu}\caA_\nu) \xi,\label{curvsuper}
\end{align}
for the non-vanishing contributions, with the indices defined in \eqref{indexform}. Remember that $F_{\mu\nu}$ and $\caA_\mu$ have been defined in section \ref{qft}. The gauge action which can be built from the differential calculus writes
\begin{equation}
S(A)=\frac14\tr(\sum_{f,g}|\caF_{f,g}|^2),\label{actsupergauge}
\end{equation}
where like in the scalar case, the sum for $f,g$ is performed over $i\xi,-\frac{i}{2}\wx_\mu,-\frac{i}{2}\wx_\mu\xi$. By simplifying the expression of \eqref{actsupergauge}, one obtains
\begin{equation*}
S(A)=\int\dd^4x\Big(\frac 14F_{\mu\nu}\star F_{\mu\nu}+\frac{\alpha^2}{4}\{\caA_\mu,\caA_\nu\}_\star^2+\frac{4}{\theta^2} \caA_\mu\star\caA_\mu\Big),
\end{equation*}
which is exactly the expression of the gauge model \eqref{acteff}.

\section{Discussion}
\label{disc}

In this section, we will discuss the interpretation of the Langmann-Szabo duality in the supergeometric context developed in the previous section. More precisely, we prove that the Langmann-Szabo duality corresponds to the grading exchange, which can now be extended to the gauge theory. To do that, we show the construction of the differential calculus of section \ref{super} (namely the Lie superalgebra $\kg$) from the Heisenberg algebra $\kh$ (see \eqref{heisenberg}) used in section \ref{lsd}.

First, we consider an extension of the complexification of $\kh$: $\tilde\kh=\gC^{10}$ with commutation relations: 
$\forall x,y,p,q\in\gC^4$, $\forall s,t,a,b\in\gC$,
\begin{equation}
[(x,p,s,a),(y,q,t,b)]=(0,0,-x\Sigma q-p\Sigma y, -i\theta x\Sigma y-i\alpha \theta p\Sigma q).\label{relext}
\end{equation}
This corresponds to provide noncommutative position coordinates with deformation parameter $\theta$ as well as noncommutative momentum coordinates with deformation parameter $\alpha\theta$. We can then see $\tilde\kh$ as a Lie subalgebra of $\algA_\theta$ for the non-graded star-commutator:
\begin{equation}
(y,q,t,b)\in\tilde\kh\mapsto -\frac{i\theta}{2} y_\mu\wx_\mu-\frac{i\theta}{2}q_\mu\wx_\mu\xi+i\theta t\xi+b\in\algA_\theta.\label{transl}
\end{equation}
The non-vanishing commutation relations of $\algA_\theta$ (for $[\fois,\fois]$ the non-graded bracket in $\algA_\theta$):
\begin{align*}
&[-\frac{i\theta}{2}\wx_\mu,-\frac{i\theta}{2}\wx_\nu]=-i\theta\Sigma_{\mu\nu},\qquad [-\frac{i\theta}{2}\wx_\mu,-\frac{i\theta}{2}\wx_\nu\xi]=-i\theta\Sigma_{\mu\nu}\xi,\\ &[-\frac{i\theta}{2}\wx_\mu\xi,-\frac{i\theta}{2}\wx_\nu\xi]=-i\alpha\theta\Sigma_{\mu\nu},
\end{align*}
are a reexpression of \eqref{relext}.

The next step is to superize $\tilde\kh$, namely to replace the bracket of two odd elements (for the grading of $\algA_\theta$) by the graded bracket of these elements in $\algA_\theta$. The modified relations hold:
\begin{equation*}
[i\theta\xi,i\theta\xi]_\star=-2\alpha\theta^2,\qquad [-\frac{i\theta}{2}\wx_\mu\xi,i\theta\xi]_\star=\alpha\theta\wx_\mu,\qquad [-\frac{i\theta}{2}\wx_\mu\xi,-\frac{i\theta}{2}\wx_\nu\xi]_\star=-\frac{\alpha}{2}\wx_\mu\wx_\nu,
\end{equation*}
where $[\fois,\fois]_\star$ is the graded bracket \eqref{grbracket} of $\algA_\theta$. We see in the last relation that $\tilde\kh$ is not closed for this graded bracket. By adding the generator $\frac{i}{2}\wx_\mu\wx_\nu$, one obtains a Lie superalgebra isomorphic to $\kg$, defined in section \ref{super}. Therefore, $\kg$ is constructed from $\kh$ by an extension and a superization.

The symplectic group $Sp(\gR^8,\omega)$ which was acting on the Heisenberg algebra $\kh$ (see \eqref{actsympl}) is no longer a symmetry group of $\kg$ (it does not preserve the graded bracket). However, the matrix $M$ in $Sp(\gR^8,\omega)$, corresponding to the Langmann-Szabo duality, still acts on $\kg$ as a linear application. By using \eqref{transl}, we obtain
\begin{equation*}
M.(\wx_\mu)=\frac{\theta}{2}\wx_\mu\xi,\qquad M.(\wx_\mu\xi)=\frac{2}{\theta}\wx_\mu,\qquad M.(\xi)=\xi,\qquad M.(\wx_\mu\wx_\nu)=\wx_\mu\wx_\nu.
\end{equation*}
Thus, the Langmann-Szabo duality corresponds to the grading exchange for superfunctions of $\algA_\theta$ which are first order polynomials in the even variable $x$. Note that, contrary to the Langmann-Szabo duality, the grading exchange can be extended to the gauge theory \eqref{actsupergauge}. In Equations \eqref{curvsuper}, we see that the grading exchange ($\mu,\nu\leftrightarrow\overline\mu,\overline\nu$) is a symmetry between the star-commutator term and the star-anticommutator one, exactly like the one noticed in the action \eqref{actgaugesym}.

Moreover, one can make two remarks. The action of the matrix $M$ is not a symmetry of $\kg$, but one can restore it as a symmetry for the scalar field theory by identifying the even and the odd part of the field, i.e. by taking $(1+\xi)\phi(x)$ as the scalar field, as well as for the gauge theory by assuming $A_\mu^0=A_\mu^1$ (see section \ref{super}). Furthermore, the generator $\frac{i}{2}\wx_\mu\wx_\nu$ can be viewed as auxiliary since it arises only from the closure of the Lie superalgebra. Consequently, the scalar and gauge actions are constructed in section \ref{super} without taking the contribution of this generator.

\medskip

To conclude, we have shown here how supergeometry could be used to interpret noncommutative field theories with harmonic term. We indeed change the point of view. Instead of adding terms (like the harmonic term) to the action in order to solve the UV-IR mixing problem, we consider the standard action but in a superized framework: we have introduced $\algA_\theta$, the deformation quantization of $\gR^{4|1}$, and constructed a differential calculus on $\algA_\theta$ thanks to a superization of the Heisenberg algebra ($\kh\mapsto\kg$). Then, we obtain the Grosse-Wulkenhaar model as well as the gauge theory with harmonic term \eqref{acteff}, as the standard actions for this superalgebra and this differential calculus (with its theory of connections). Moreover, the Langmann-Szabo duality for the scalar theory corresponds to the grading exchange in $\algA_\theta$ and can be extended to the gauge theory. This supergeometric interpretation of the removing of UV-IR mixing is interesting as potentially generalizable to other noncommutative spaces also suffering from this mixing \cite{Gayral:2005af}.

\acknowledgments
The author thanks the organizers of the {\it Corfu Summer Institute on Elementary Particles and Physics 2010} for their invitation.

\bibliographystyle{utcaps}
\bibliography{biblio-these,biblio-perso,biblio-recents}

\providecommand{\href}[2]{#2}\begingroup\raggedright\begin{thebibliography}{10}

\bibitem{Wulkenhaar:2006si}
R.~Wulkenhaar, ``{Field theories on deformed spaces},''
\href{http://dx.doi.org/10.1016/j.geomphys.2005.04.019}{{\em J. Geom. Phys.}
  {\bf 56} (2006)  108--141}.

\bibitem{Connes:1994}
A.~Connes, {\em {Noncommutative Geometry}}.
\newblock Academic Press, San Diego, New York, London, 1994.

\bibitem{Doplicher:1994tu}
S.~Doplicher, K.~Fredenhagen, and J.~E. Roberts, ``{The Quantum structure of
  space-time at the Planck scale and quantum fields},''
  \href{http://dx.doi.org/10.1007/BF02104515}{{\em Commun. Math. Phys.} {\bf
  172} (1995)  187--220},
\href{http://arxiv.org/abs/hep-th/0303037}{{\tt arXiv:hep-th/0303037}}.

\bibitem{Chamseddine:2006ep}
A.~H. Chamseddine, A.~Connes, and M.~Marcolli, ``{Gravity and the standard
  model with neutrino mixing},'' {\em Adv. Theor. Math. Phys.} {\bf 11} (2007)
  991--1089,
\href{http://arxiv.org/abs/hep-th/0610241}{{\tt arXiv:hep-th/0610241}}.

\bibitem{Freidel:2006}
L.~Freidel and E.~R. Livine, ``{3D quantum gravity and effective noncommutative
  quantum field theory},'' {\em Phys. Rev. Lett.} {\bf 96} (2006)  221301.

\bibitem{Seiberg:1999vs}
N.~Seiberg and E.~Witten, ``{String theory and noncommutative geometry},'' {\em
  JHEP} {\bf 09} (1999)  032,
\href{http://arxiv.org/abs/hep-th/9908142}{{\tt arXiv:hep-th/9908142}}.

\bibitem{Connes:1997cr}
A.~Connes, M.~R. Douglas, and A.~S. Schwarz, ``{Noncommutative geometry and
  matrix theory: Compactification on tori},'' {\em JHEP} {\bf 02} (1998)  003,
\href{http://arxiv.org/abs/hep-th/9711162}{{\tt arXiv:hep-th/9711162}}.

\bibitem{Hellerman:2001rj}
S.~Hellerman and M.~Van~Raamsdonk, ``{Quantum Hall physics equals
  noncommutative field theory},'' {\em JHEP} {\bf 10} (2001)  039,
\href{http://arxiv.org/abs/hep-th/0103179}{{\tt arXiv:hep-th/0103179}}.

\bibitem{Minwalla:1999px}
S.~Minwalla, M.~Van~Raamsdonk, and N.~Seiberg, ``{Noncommutative perturbative
  dynamics},'' {\em JHEP} {\bf 02} (2000)  020,
\href{http://arxiv.org/abs/hep-th/9912072}{{\tt arXiv:hep-th/9912072}}.

\bibitem{Grosse:2004yu}
H.~Grosse and R.~Wulkenhaar, ``{Renormalisation of phi**4 theory on
  noncommutative R**4 in the matrix base},''
  \href{http://dx.doi.org/10.1007/s00220-004-1285-2}{{\em Commun. Math. Phys.}
  {\bf 256} (2005)  305--374},
\href{http://arxiv.org/abs/hep-th/0401128}{{\tt arXiv:hep-th/0401128}}.

\bibitem{deGoursac:2007gq}
A.~de~Goursac, J.-C. Wallet, and R.~Wulkenhaar, ``{Noncommutative induced gauge
  theory},'' \href{http://dx.doi.org/10.1140/epjc/s10052-007-0335-2}{{\em Eur.
  Phys. J.} {\bf C51} (2007)  977--987},
\href{http://arxiv.org/abs/hep-th/0703075}{{\tt arXiv:hep-th/0703075}}.

\bibitem{deGoursac:2009gh}
A.~de~Goursac, {\em {Noncommutative geometry, gauge theory and
  renormalization}}.
\newblock Verlag Dr. Muller, Saarbrucken, 2010.

\bibitem{deGoursac:2008bd}
A.~de~Goursac, T.~Masson, and J.-C. Wallet, ``{Noncommutative
  $\varepsilon$-graded connections},'' {\em to appear in J. Noncommut. Geom.}
  ,
\href{http://arxiv.org/abs/0811.3567}{{\tt arXiv:0811.3567 [math-ph]}}.

\bibitem{deGoursac:2010zb}
A.~de~Goursac, ``{On the origin of the harmonic term in noncommutative quantum
  field theory},'' \href{http://dx.doi.org/10.3842/SIGMA.2010.048}{{\em SIGMA}
  {\bf 6} (2010)  048},
\href{http://arxiv.org/abs/1003.5788}{{\tt arXiv:1003.5788 [math-ph]}}.

\bibitem{Bieliavsky:2010su}
P.~Bieliavsky, A.~de~Goursac, and G.~Tuynmann, ``{Deformation quantization for
  Heisenberg supergroup},'' \href{http://arxiv.org/abs/1011.2370}{{\tt
  arXiv:1011.2370 [math.QA]}}.

\bibitem{Buric:2009ss}
M.~Buric and M.~Wohlgenannt, ``{Geometry of the Grosse-Wulkenhaar Model},''
  \href{http://dx.doi.org/10.1007/JHEP03(2010)053}{{\em JHEP} {\bf 03} (2010)
  053},
\href{http://arxiv.org/abs/0902.3408}{{\tt arXiv:0902.3408 [hep-th]}}.

\bibitem{Buric:2010xs}
M.~Buric, H.~Grosse, and J.~Madore, ``{Gauge fields on noncommutative
  geometries with curvature},''
  \href{http://dx.doi.org/10.1007/JHEP07(2010)010}{{\em JHEP} {\bf 07} (2010)
  010},
\href{http://arxiv.org/abs/1003.2284}{{\tt arXiv:1003.2284 [hep-th]}}.

\bibitem{Fischer:2008dq}
A.~Fischer and R.~J. Szabo, ``{Duality covariant quantum field theory on
  noncommutative Minkowski space},''
  \href{http://dx.doi.org/10.1088/1126-6708/2009/02/031}{{\em JHEP} {\bf 02}
  (2009)  031},
\href{http://arxiv.org/abs/0810.1195}{{\tt arXiv:0810.1195 [hep-th]}}.

\bibitem{Zahn:2010yt}
J.~Zahn, ``{Divergences in quantum field theory on the noncommutative
  two-dimensional Minkowski space with Grosse-Wulkenhaar potential},''
\href{http://arxiv.org/abs/1005.0541}{{\tt arXiv:1005.0541 [hep-th]}}.

\bibitem{Fischer:2010zg}
A.~Fischer and R.~J. Szabo, ``{UV/IR duality in noncommutative quantum field
  theory},''
\href{http://arxiv.org/abs/1001.3776}{{\tt arXiv:1001.3776 [hep-th]}}.

\bibitem{Bayen:1978}
F.~Bayen, M.~Flato, C.~Fronsdal, A.~Lichnerowicz, and D.~Sternheimer,
  ``{Deformation theory and quantization},'' {\em Ann. Phys.} {\bf 11} (1978)
  61--151.

\bibitem{Tanasa:2007xa}
A.~Tanasa and F.~Vignes-Tourneret, ``{Hopf algebra of non-commutative field
  theory},'' {\em J. Noncommut. Geom.} {\bf 2} (2008)  125--139,
\href{http://arxiv.org/abs/0707.4143}{{\tt arXiv:0707.4143 [math-ph]}}.

\bibitem{Tanasa:2009hb}
A.~Tanasa and D.~Kreimer, ``{Combinatorial Dyson-Schwinger equations in
  noncommutative field theory},''
\href{http://arxiv.org/abs/0907.2182}{{\tt arXiv:0907.2182 [hep-th]}}.

\bibitem{Disertori:2006nq}
M.~Disertori, R.~Gurau, J.~Magnen, and V.~Rivasseau, ``{Vanishing of beta
  function of non commutative phi(4)**4 theory to all orders},''
  \href{http://dx.doi.org/10.1016/j.physletb.2007.04.007}{{\em Phys. Lett.}
  {\bf B649} (2007)  95--102},
\href{http://arxiv.org/abs/hep-th/0612251}{{\tt arXiv:hep-th/0612251}}.

\bibitem{deGoursac:2007uv}
A.~de~Goursac, A.~Tanasa, and J.~C. Wallet, ``{Vacuum configurations for
  renormalizable non-commutative scalar models},''
  \href{http://dx.doi.org/10.1140/epjc/s10052-007-0465-6}{{\em Eur. Phys. J.}
  {\bf C53} (2008)  459--466},
\href{http://arxiv.org/abs/0709.3950}{{\tt arXiv:0709.3950 [hep-th]}}.

\bibitem{deGoursac:2009fm}
A.~de~Goursac and J.-C. Wallet, ``{Symmetries of noncommutative scalar field
  theory},'' {\em J. Phys. A: Math. Theor.} {\bf 44} (2011)  055401,
\href{http://arxiv.org/abs/0911.2645}{{\tt arXiv:0911.2645 [math-ph]}}.

\bibitem{Gurau:2008vd}
R.~Gurau, J.~Magnen, V.~Rivasseau, and A.~Tanasa, ``{A translation-invariant
  renormalizable non-commutative scalar model},''
  \href{http://dx.doi.org/10.1007/s00220-008-0658-3}{{\em Commun. Math. Phys.}
  {\bf 287} (2009)  275--290},
\href{http://arxiv.org/abs/0802.0791}{{\tt arXiv:0802.0791 [math-ph]}}.

\bibitem{Tanasa:2010fk}
A.~Tanasa, ``{Translation-Invariant Noncommutative Renormalization},'' {\em
  SIGMA} {\bf 6} (2010)  047,
\href{http://arxiv.org/abs/1003.4877}{{\tt arXiv:1003.4877 [hep-th]}}.

\bibitem{Cagnache:2008tz}
E.~Cagnache, T.~Masson, and J.-C. Wallet, ``{Noncommutative Yang-Mills-Higgs
  actions from derivation- based differential calculus},'' {\em J. Noncommut.
  Geom.} {\bf 5} (2011)  39,
\href{http://arxiv.org/abs/0804.3061}{{\tt arXiv:0804.3061 [hep-th]}}.

\bibitem{Grosse:2007dm}
H.~Grosse and M.~Wohlgenannt, ``{Induced Gauge Theory on a Noncommutative
  Space},'' \href{http://dx.doi.org/10.1140/epjc/s10052-007-0369-5}{{\em Eur.
  Phys. J.} {\bf C52} (2007)  435--450},
\href{http://arxiv.org/abs/hep-th/0703169}{{\tt arXiv:hep-th/0703169}}.

\bibitem{deGoursac:2008rb}
A.~de~Goursac, J.-C. Wallet, and R.~Wulkenhaar, ``{On the vacuum states for
  noncommutative gauge theory},''
  \href{http://dx.doi.org/10.1140/epjc/s10052-008-0652-0}{{\em Eur. Phys. J.}
  {\bf C56} (2008)  293--304},
\href{http://arxiv.org/abs/0803.3035}{{\tt arXiv:0803.3035 [hep-th]}}.

\bibitem{Blaschke:2007vc}
D.~N. Blaschke, H.~Grosse, and M.~Schweda, ``{Non-commutative U(1) gauge theory
  on R**4(Theta) with oscillator term and BRST symmetry},''
  \href{http://dx.doi.org/10.1209/0295-5075/79/61002}{{\em Europhys. Lett.}
  {\bf 79} (2007)  61002},
\href{http://arxiv.org/abs/0705.4205}{{\tt arXiv:0705.4205 [hep-th]}}.

\bibitem{Blaschke:2009aw}
D.~N. Blaschke, H.~Grosse, E.~Kronberger, M.~Schweda, and M.~Wohlgenannt,
  ``{Loop Calculations for the Non-Commutative U(1) Gauge Field Model with
  Oscillator Term},''
\href{http://arxiv.org/abs/0912.3642}{{\tt arXiv:0912.3642 [hep-th]}}.

\bibitem{Langmann:2002cc}
E.~Langmann and R.~J. Szabo, ``{Duality in scalar field theory on
  noncommutative phase spaces},''
  \href{http://dx.doi.org/10.1016/S0370-2693(02)01650-7}{{\em Phys. Lett.} {\bf
  B533} (2002)  168--177},
\href{http://arxiv.org/abs/hep-th/0202039}{{\tt arXiv:hep-th/0202039}}.

\bibitem{Bieliavsky:2008qy}
P.~Bieliavsky, R.~Gurau, and V.~Rivasseau, ``{Non Commutative Field Theory on
  Rank One Symmetric Spaces},'' {\em J. Noncommut. Geom.} {\bf 3} (2009)
  99--123,
\href{http://arxiv.org/abs/0806.4255}{{\tt arXiv:0806.4255 [hep-th]}}.

\bibitem{DuboisViolette:1988cr}
M.~Dubois-Violette, ``{D{\'e}rivations et calcul differentiel non
  commutatif},'' {\em C.R. Acad. Sci. Paris, S\'erie} {\bf I} (1988)  307:
  403--408.

\bibitem{Gayral:2005af}
V.~Gayral, ``{Non compact isospectral deformations and quantum field theory},''
\href{http://arxiv.org/abs/hep-th/0507208}{{\tt arXiv:hep-th/0507208}}.

\end{thebibliography}\endgroup

\end{document}